\documentclass[12pt]{article}
\usepackage{epsf}

\newcommand{\beq}{\begin{equation}}
\newcommand{\eeq}{\end{equation}}
\newcommand{\beqcol}{\begin{array}{rcl}}
\newcommand{\eeqcol}{\end{array}}
\renewcommand{\th}{\theta}

\begin{document}

{\thispagestyle{empty}

\begin{flushright}  
KCL-MTH-98-64 \\
December 1998 
\end{flushright}
\vfill
\begin{center}
{\Large 
%{\obeylines
Exact two-particle Matrix Elements in S-Matrix Preserving \\[3mm]
Deformation of Integrable QFTs}
\\
\vspace{2cm}
Mathias Pillin\footnote{e-mail: map@mth.kcl.ac.uk}    \\
\vspace{1cm}
Department of Mathematics, King's College \\
Strand, London WC2R 2LS, U.K.   \\ 
\medskip

\vfill
{\bf Abstract}
\end{center}
\begin{quote}
In a recent paper it was shown that the response of an 
integrable QFT under variation of the Unruh temperature 
can be computed from a S-matrix preserving deformation
of the form factor approach. We give explicit expressions 
for the deformed two-particle formfactors for various 
integrable models: The Sine-Gordon and $SU(2)$ Thirring model, 
several perturbed minimal CFTs and the real coupling affine Toda 
series. A uniform pattern is found to emerge when both  
the S-matrix and the deformed form factors are expressed 
in terms Barnes' multi-periodic functions.
\end{quote}

\eject
}

\setcounter{page}{1}

{\bf 1. Introduction}

\bigskip
\bigskip

The formfactor approach provides a powerful and rigid technique to 
construct and to solve a large class of 1+1 dimensional quantum field 
theories (QFTs). Its computational rigidity however also provides 
an obstacle when one tries to apply it to non-standard situations, 
even when the problem is essentially of a quantum field theoretical 
nature. Computing the response of a QFT under a variation of the 
Unruh temperature is such a problem and in a recent paper \cite{MAX1} 
a generalized formfactor framework was shown to be able to cope 
with the problem, while retaining the virtues of the usual formfactor 
approach \cite{KW,SMIR}. In essence each integrable QFT admits a 
one-parameter deformation that preserves the bootstrap S-matrix. 
The deformation parameter $\beta$ plays the role of an inverse 
Unruh temperature, $\beta = 2\pi$ being the QFT value.

The purpose of this letter is to prepare the ground for a 
more detailed investigation of these systems by computing the 
deformed two-particle formfactors, i.e. the non-perturbative 
two-particle matrix elements of a local operator, for many of the commonly
considered models. These two-particle matrix elements 
$F^m(\th_2 - \th_1)$ are of particular interest because 
typically a factor $\prod_{k > l}^n F^m(\th_k - \th_l)$ 
appears in a successful Ansatz for the n-particle formfactors 
\cite{SMIR}, 
a feature that seems to be preserved under deformation \cite{MAX1}. 
Splitting off this factor results in a recurvise problem
for the remainder in an often much simpler function space 
(ideally one of polynomials, see e.g. \cite{MAP, OOTA}).    

Explicitly the defining relations for the minimal 
two-particle deformed form factor are  
\beq
\beqcol
F^{m} (\theta ) & = & S (\theta ) F^m(-\theta ) , \\
F^{m} (\theta + i \beta ) &=& F^m(-\theta ) , 
\eeqcol
\label{watson}
\eeq

where $S(\theta)$ denotes the exact two-particle S-matrix. 
We will consider a version of (\ref{watson}) with indices 
when needed.

The term ``minimal'' refers to the condition that the solution 
searched for should have no poles and zeros in the strip $0 <
{\rm Im}\,\theta < \beta/2$ and just a simple zero at $\th =0$; the 
kinematical poles can readily
be incorporated by multiplying with a suitable $i\beta$-periodic 
symmetric function. In the limit $\beta= 2 \pi$ the equations 
(\ref{watson}) are known as Watson's equations; the novel features
of the $\beta \neq 2\pi$ case can already be anticipated by noting 
that e.g.~for a $2\pi i$-periodic S-matrix, the period of the S-matrix 
and the (cyclicity) period of the form factors no longer coincide. 
Likewise the kinematical poles appear at relative rapidities
$\pm i\pi$ and {\em not} at $\pm i\beta/2$.

The purpose of the present letter is to present solutions of 
(\ref{watson}) for generic $\beta$ for many of the commonly 
considered integrable QFTs: The Sine-Gordon and $SU(2)$ Thirring model, 
perturbed minimal CFTs of the $A_{2N}^{(2)}$ series as well as
the simply-laced real coupling affine Toda theories. 
It turns out that Barnes' multi-periodic functions \cite{BARNES} 
provide
the proper tool to find and describe the solutions. Moreover  
a uniform pattern is found to emerge when both  
the S-matrix and the deformed form factors are expressed 
in terms of Barnes' functions.

\bigskip
\bigskip

%%%%%%%%%%%%%%%%%%%%%%%%%%%%%%%%%%%%%%%%%%%%%%%%%%%%%%%%%%%%%%%%%%%%%

{\bf 2. Barnes' multi-periodic functions} 

\bigskip

We are going to recall several facts about multi-periodic 
functions. The mathematical theory of these functions 
had been developed by Barnes \cite{BARNES} a long time 
ago. We use the conventions of \cite{BARNES} and borrow results 
from \cite{JM} where additional properties of the functions 
in question may be found. 
\medskip

Let ${\underline{\omega}}= ( \omega_1, \omega_2, \ldots , \omega_r )$
denote the vector of periods, where ${\rm{Re}}\,\omega_i>0$. 
Set $ |{\underline{\omega}}| = \omega_1 + \ldots + \omega_r $, and let 
${\underline{\omega}}(i)$ be the vector of periods with the 
period $\omega_i$ omitted. Following \cite{BARNES,WW} we define 
the multi-periodic zeta-function $\zeta_{r}( s,x | {\underline{\omega}} )$ via 
its contour integral representation

\beq
\zeta_{r}( s,x | {\underline{\omega}} ) = - { {\Gamma (1-s)}\over{2 \pi i}} 
\int_{C_H}   { { \exp( -x t) (-t)^{s-1} } \over 
               {\prod_{i=1}^r (1- \exp ( - \omega_i t ) ) }} 
   {\rm{d}}t .
\label{zeta-def}
\eeq

The integration curve $C_H$ is of Hankel type (see e.g. \cite{WW}, 
chapter 12), which means that after a deformation of the contour 
we integrate from infinity back to a small circle (counterclockwise) 
around the 
origin and then back to positive infinity. It was shown in 
\cite{BARNES} that the integral in (\ref{zeta-def}) can be 
asymptotically written in terms of infinite sums, hence 
resembling the Hurwitz form of $\zeta$. 

\medskip

Further we introduce the multiple $\Gamma$-function via a 
derivative of the function $\zeta_r$ in the integral 
representation (\ref{zeta-def}), i.e.
 
\beq
\log \Gamma_r(x | {\underline{\omega}} ) = { {\partial}\over{\partial s}}
 \left. \zeta_{r}( s ,x | {\underline{\omega}} )\right|_{s=0}
 =  { {1 }\over{2 \pi i}} 
\int_{C_H}   { { \exp( -x t) ( \log ( -t) - \gamma) } \over 
               {\prod_{i=1}^r (1- \exp ( - \omega_i t ) ) }} 
   { { {\rm{d}}t }\over{t}},
\label{gamma-def}
\eeq

where $\gamma$ denotes the Euler constant. The functions $\Gamma_r$ 
in (\ref{gamma-def}) are meromorphic with poles at 
$x= n_1 \omega_1+ n_2 \omega_2 + \ldots + n_r \omega_r$, for 
$n_i$ being non-positive integers.

\smallskip

In order 
to compare $\Gamma_1$ with the standard $\Gamma$-function, 
the latter being $1$ periodic by definition, we note that

\beq
\Gamma_1( x | \omega_1 ) = \omega_1^{x/\omega_1 -1/2}
 \; \Gamma(x/\omega_1 ) / \sqrt{2 \pi}  .
\label{gamma-1}
\eeq

The multi-periodicity of the $\Gamma$-functions defined in 
(\ref{gamma-def}) is reflected in the following relation. 

\beq
{  { \Gamma_r ( x+ \omega_i | {\underline{\omega}}) } \over 
 { \Gamma_r ( x | {\underline{\omega}}) }} = 
{ {1} \over { \Gamma_{r-1} ( x | {\underline{\omega}}(i)   ) }} .
\label{gamma-period}
\eeq

\medskip

Finally we define the multiperiodic sine-function by 
\cite{BARNES}

\beq
S_{r} ( x | {\underline{\omega}}) = \Gamma_{r} ( x | {\underline{\omega}})^{-1} 
 \Gamma_{r} ( | {\underline{\omega}} | - x | {\underline{\omega}})^{(-1)^r}.
\label{sinus-def}
\eeq

As a consequence of (\ref{gamma-1}) the function $S_1$ is 
related to the standard sine-function by

\beq
S_1 ( x| \omega_1 ) = 2 \sin \left( { { \pi x}\over{\omega_1}} \right).
\label{sinus-1}
\eeq

The periodicity of the multi-sine functions can be derived 
from (\ref{sinus-def}) and (\ref{gamma-period}) as 

\beq
{  { S_r ( x+ \omega_i | {\underline{\omega}}) } \over 
 { S_r ( x | {\underline{\omega}}) }} = 
{ {1} \over { S_{r-1} ( x | {\underline{\omega}}(i)   ) }} .
\label{sinus-period}
\eeq  

We mention for later purposes that $S_2$ is meromorphic 
with poles at $x=n_1 \omega_1 + n_2 \omega_2$, with $m_1,m_2$ 
positive integers, and zeros in $x= m_1 \omega_1 + m_2 \omega_2$, 
for $m_1, m_2$ non-positive integers. Moreover, the function 
$S_3$ has no poles but zeros in $x= m_1 \omega_1 + m_2 \omega_2 
+ m_3 \omega_3$, for $m_i \in {\bf{Z}}$.

\smallskip
 
These definitions are sufficient for the purposes 
of this letter. In the appendix we present an integral 
formula for $S_r$ and two infinite product expansions for $S_2$ 
in order to give the reader the possibility to compare 
our results with related expressions in other papers.

\bigskip
\bigskip

%%%%%%%%%%%%%%%%%%%%%%%%%%%%%%%%%%%%%%%%%%%%%%%%%%%%%%%%%%%%%%%%%%%%%%
{\bf 3. The deformed Sine-Gordon model} 

\bigskip

For a detailed treatment of the undeformed Sine-Gordon model and 
its form factors we refer to \cite{SMIR,KAROWSKI}. The $S$-matrix 
in this model contains a scalar part $S_0(\th)$, which is expressible 
in terms of $S_2$-functions. We denote by $\xi$ the non-perturbative 
coupling constant. The deformed two-particle problem (\ref{watson})
with this S-matrix was incidentally already considered in the physically 
completely distinct context of the $XXZ$-model in the gapless regime
\cite{JM}; our solution is related to that in \cite{JM} by a suitable 
redefinition. Explicitly

\beq
S_0(\theta) = { 
   { S_2(-i \theta | 2 \pi , \xi ) S_2(\pi + i \theta | 2 \pi , \xi ) }
  \over
   { S_2(i \theta | 2 \pi , \xi ) S_2(\pi - i \theta | 2 \pi , \xi ) }} 
= - \exp \left( -i \int_{0}^{\infty} 
{ { \sin(\theta t) \sinh \left( { {\pi-\xi}\over{2}} t \right) }
  \over
  { t \cosh \left( { { \pi t}\over{2}} \right) 
      \sinh \left( { {\xi t}\over{2}} \right) }} {\rm{d}}t \right).
\label{S-sine}
\eeq 

It can be shown using the integral representation of the diperiodic 
sine-functions (\ref{sinus-int}) that the first part of the equation 
actually reproduces the integral in (\ref{S-sine}). 
The unitarity and crossing relations for all the $S$-matrix elements 
in the sine-Gordon model are readily verified from (\ref{S-sine}) 
when taking into account the periodicity equation (\ref{sinus-period}). 

\medskip

The functional equations to be satisfied by the two-particle 
deformed formfactor in the present model are

\beq
F^m_{SG}(\theta) = S_0 ( \theta ) F^m_{SG}( -\theta), \qquad
F^m_{SG}(\theta + i \beta ) = F^m_{SG}(-\theta ). 
\label{watson-sinus}
\eeq

Up to functions which are both even and $i \beta$-periodic, and 
a normalization, the unique solution to (\ref{watson-sinus}) 
with the proper analyticity properties (mentioned in the 
introduction) is given by

\beq
F^m_{SG} (\theta ) = { 
{ S_3( -i \theta | 2 \pi , \beta, \xi )  
   S_3( \beta + i \theta | 2 \pi , \beta, \xi )}
    \over 
{ S_3( \pi -i \theta | 2 \pi , \beta, \xi )  
   S_3( \pi + \beta + i \theta | 2 \pi , \beta, \xi )}}.
\label{f-sinus}
\eeq

Already at this point we observe a feature which turns out to 
be generic for the cases to be considered in this paper: The 
two particle form factors possess the same structure as the 
$S$-matrix of the model, but the number of periods is increased 
by one. 

In the limit $\beta \to 2 \pi$ it can be verified using the
integral representation (\ref{sinus-int}) that the expression 
(\ref{f-sinus}) turns into the corresponding object in 
\cite{SMIR}.

\bigskip
\bigskip

%%%%%%%%%%%%%%%%%%%%%%%%%%%%%%%%%%%%%%%%%%%%%%%%%%%%%%%%%%%%%%%%%%%%%%%%
{\bf 3. The deformed $SU(2)$ invariant Thirring model}

\bigskip

We may again refer to \cite{SMIR} for a discussion of the 
standard $SU(2)$ invariant Thirring model. The scalar part 
of the $S$-matrix is given by the ratio of 
$\Gamma$-function, which we rewrite in terms 
of $\Gamma_1$, using the relation (\ref{gamma-1}) 

\beq
S_0^{TM}( \theta ) = { 
{ \Gamma \left( { {1}\over {2}} + { {\theta}\over{2 \pi i}} \right)
  \Gamma \left(  - { {\theta}\over{2 \pi i}} \right) }
   \over
{ \Gamma \left( { {1}\over {2}} - { {\theta}\over{2 \pi i}} \right)
  \Gamma \left(   { {\theta}\over{2 \pi i}} \right) } 
}
   =
 { { \Gamma_1 ( \pi -i \theta | 2\pi) \Gamma_1 (i \theta | 2\pi ) }
   \over
   { \Gamma_1 ( \pi + i \theta | 2\pi) \Gamma_1 (- i \theta | 2\pi ) } 
  }.
\label{s-thirring}
\eeq

In the light of the previous section we should now expect the 
two-particle formfactor in the present case, 
$F^m_{TM} (\theta ) $, to be built from $\Gamma_2$ functions. Indeed 
the functional equations to be satisfied by $F^m_{TM} (\theta ) $ 
are equivalent to (\ref{watson-sinus}) and the solution is found to be

\beq
F^m_{TM} (\theta ) = {
{ \Gamma_2 ( \pi - i \theta | 2 \pi, \beta ) 
     \Gamma_2 (\pi + \beta + i \theta | 2 \pi , \beta ) }
   \over 
{ \Gamma_2 (  - i \theta | 2 \pi, \beta ) 
     \Gamma_2 ( \beta + i \theta | 2 \pi , \beta ) }
}.
\label{f-thirring}
\eeq

This expression has again the required analyticity structure, 
mentioned in the introduction and turns into the proper 
undeformed expression \cite{SMIR} for $\beta \to 2 \pi$.

It is intriguing to compare the structure of (\ref{f-thirring}) 
with the $S$-matrix (\ref{s-thirring}) and also with the 
two-particle formfactor in the sine-Gordon model 
(\ref{f-sinus}). 

Note that in the limit $\beta \to 2 \pi $ the $\Gamma_2$ functions 
in (\ref{f-thirring}) have two equal periods. It was shown in 
\cite{BARNES} that this does not allow (apart from infinite 
product expansions) to rewrite (\ref{f-thirring}) in simple terms 
of functions with less periods.

\bigskip
\bigskip

%%%%%%%%%%%%%%%%%%%%%%%%%%%%%%%%%%%%%%%%%%%%%%%%%%%%%%%%%%%%%%%%%%%%
{\bf 4. Affine Toda theories and perturbed minimal models}

\bigskip

In this section we compute the two-particle 
form factors for the deformed affine Toda models of 
type ADE and of the deformed perturbed minimal models 
of the $A_{2N}^{(2)}$ series. 

As it was shown in \cite{SMIR2} the latter models can in 
the standard (undeformed) 
context be understood as rational reductions of the breather 
sectors in the Sine-Gordon model. From the point of view 
of the $S$-matrices and the two-particle formfactors, this 
process can be easily reversed. It is then straightforward 
in the deformed case to be described below to reconstruct 
the deformed two-particle breather formfactors from 
our results.

In this section, we first have to review a few things 
about the $S$-matrices of the affine Toda theories and 
the perturbed minimal models resp., and to introduce 
some notations. 

It is known \cite{BCDS} that the basic constituent of the $S$-matrices 
for the models under consideration is the following object.

\beq
( x)_{\theta} = { { ( x)_{+} } \over{(-x)_{+}}}, \qquad 
(x)_{+} = \sinh { {1}\over{2}} \left( \theta + { {i \pi}\over{h}} x 
                    \right), 
\label{konv-1}
\eeq

where $h$ denotes the Coxeter number of the underlying Lie 
algebra. The building block of the $S$-matrices is given by 

\beq
\langle x \rangle_{\theta} = { {\langle x \rangle_{+}} \over 
            {\langle - x \rangle_{+} }}, 
 \qquad
\langle x \rangle_{+} = \left[   \begin{array}{ll} 
      (x-1)_{+} (x+1)_{+} & \mbox{for perturbed conformal} \\
      { { (x-1)_{+} (x+1)_{+} } \over { (x-1+B)_{+} (x+1-B)_{+}}}
                  & \mbox{for affine Toda models}
                    \end{array}
         \right.
\label{konv-2}
\eeq
                  
$B$ is the non-perturbative coupling of the affine Toda models \cite{BCDS}.

The $S$-matrices for both cases to be considered in this section can 
the be written as a product, where $x$ takes values in a certain 
set $A_{ab}$, which is specified by Lie algebraic data, or 
using Weyl group techniques \cite{DOREY}, where $m_{ab}(p)$ will 
denote an exponent obtained via this method.

\beq
S_{ab}(\theta ) = \prod_{x \in A_{ab} }  \langle x \rangle_{\theta} 
       = \prod_{p=1}^{h-1} (\langle x \rangle_{\theta} )^{m_{ab}(p)}.
\label{s-toda}
\eeq

\medskip

We have now collected enough material to compute the deformed 
two-particle form factors in the present case. The functional 
equations to be solved are 

\beq
F^m_{ab} (\theta) = S_{ab} (\theta )\; F^m_{ba}(-\theta), 
\qquad
F^m_{ab}(\theta+ i \beta ) = F^{m}_{ba}(-\theta ).
\label{toda-watson}
\eeq 
The indices stand for the particle species. 

In what follows we adopt the notation used in \cite{OOTA} 
for the construction of the two-particle 
formfactors in the standard context. However, we would 
like to mention that the objects used in that paper are 
defined by infinite product expansions of $\Gamma$ 
functions which are not convergent. 

\medskip

The structure of the $S$-matrices suggests to work with 
$S_2$-functions to solve (\ref{toda-watson}). To that 
end, we define

\beq
g_{x}( \theta ) = S_2 ( -i \theta + { {\pi}\over{h}} x | 2\pi, \beta )
\label{g-def}
\eeq

Obviously it holds that $g_{x}(\theta + i {{\pi}\over{h}} y ) = 
g_{x+y} ( \theta ) $. Using the periodicity of the $S_2$-function 
(\ref{sinus-period}) we verify the following identities

\beq
g_{x}( \theta + i \beta) ={  { g_{x}(\theta ) } \over { (-2 i) 
  \sinh { {1}\over{2}} \left( \theta + i { {\pi}\over{h}} x \right) }}, 
\qquad
g_{x}( \theta + i 2 \pi) ={  { g_{x}(\theta ) } \over { (-2 i) 
  \sinh { {\pi}\over{\beta}} \left( \theta + i { {\pi}\over{h}} x \right) }}.
\label{identi-1}
\eeq

The definition of the function $S_2$ in (\ref{sinus-def}) allows the 
identification

\beq
S_2 ( \omega_1 + \omega_2 -x | \omega_1 , \omega_2 ) = { {1} \over
   {S_2 (x | \omega_1 , \omega_2 )}} ,
\label{s2-identi}
\eeq

which we need in order to verify the following

\beq
g_{x} ( i \pi - \theta )  = { {1}\over { g_{3h-x}(\theta )}} 
= -2 i\; \sinh { {1}\over{2}} \left( \theta + i { {\pi}\over{h} } (h-x) 
                               \right)   
        {{1}       \over
       {g_{h-x} (\theta ) }}.
\label{identi-2}
\eeq

As in the case of the $S$-matrices we can now construct a 
building block for the two-particle formfactors 

\beq
G_{x} (\theta) = \left[ \begin{array}{ll} 
       g_{x-1}(\theta) g_{x+1} (\theta) & \mbox{for perturbed conformal}\\
       { { g_{x-1}(\theta) g_{x+1}(\theta)}\over
         { g_{x-1+B}(\theta) g_{x+1-B}}(\theta)} 
&\mbox{for affine Toda models }
       \end{array}
   \right.
\label{toda-blocks}
\eeq
  
Notice that using (\ref{sin-prod-1}) we get a convergent expansion 
of these blocks in terms of $\Gamma$ functions.

Using the identities (\ref{identi-1}) and (\ref{identi-2}) we 
can easily verify that the following expression satisfies 
the functional equations (\ref{toda-watson}).

\beq
F_{ab}^m(\theta) = \prod_{x\in A_{ab}} f_{x}(\theta), \qquad 
f_{x}(\theta ) = { { G_{x}(\theta )}\over{G_{2 h-x}(\theta ) }}
\label{toda-loesung}
\eeq

The analyicity of this solution is the one we required 
in the introduction.

Comparing this solution (written in terms of $S_2$-functions) 
with the corresponding results in the Sine-Gordon and Thirring 
models, respectively, one sees that their structure is 
in principle equivalent.

\medskip

The expressions (\ref{toda-loesung}) also faciliate the derivation
of additional functional relations for the two-particle formfactors 
which in the undeformed case allow one to reduce the construction of 
n-particle formfactors to a polynomial problem \cite{MAP,OOTA}. 
For the Sinh-Gordon model a similar identity was shown to have 
essentially the same consequences in the deformed case \cite{MAX1}. 
To generalise this identity to other affine Toda models  
we introduce another bit of notation

\beq
[x]^{\beta}_{+} = { 
  { \sinh  {{\pi}\over{\beta}} \left( \theta + i {{\pi}\over{h}} (x-1) \right)
    \sinh { {\pi}\over{\beta}} \left( \theta + i {{\pi}\over{h}} (x+1) \right)}
 \over
 {\sinh { {\pi}\over{\beta}} \left( \theta + i {{\pi}\over{h}} (x-1+B) \right) 
 \sinh{ {\pi}\over{\beta}} \left( \theta + i {{\pi}\over{h}} (x+1-B) \right)
}}.
\label{konv3}
\eeq

This is the expression we need for the affine Toda models. According 
to (\ref{konv-2}) the corresponding expression for the case of perturbed 
minimal models is obtained by setting the denominator in 
(\ref{konv3}) equal to one. 

The first identity stems from the crossing symmetry of the 
$S$-matrix and reads 

\beq
F^m_{\bar{a}b}(\theta+ i \pi ) F^m_{ab}(\theta) = \prod_{x\in A_{ab}} 
[x]^{\beta}_{+}. 
\label{tralala1}
\eeq

The second identity is a consequence of the boostrap equations 
of the $S$-matrix with $\theta^{c}_{ab}$, $\bar{\theta}^{c}_{ab}=
\pi - \theta^{c}_{ab}$ denoting the 
fusion angles \cite{BCDS}

\beq
\lambda^c_{ab;d}(\theta)^{-1} = 
{ F^{{\rm min}}_{ad} (\theta+ i \bar{\theta}^b_{ac} ) 
  F^{{\rm min}}_{bd}(\theta- i \bar{\theta}^a_{bc} ) \over 
{F^{{\rm min}}_{\bar{c} d} (\theta) }} = 
\prod_{p=0}^{\bar{u}^b_{ac}} 
     ( [ \bar{u}^b_{ac}- p ]^{\beta}_{+})^{m_{ad}(p)} 
\prod_{p=0}^{\bar{u}^a_{bc}-1} 
     ( [ p -\bar{u}^a_{bc} ]^{\beta}_{+})^{m_{bd}(p)} .
\label{lamdef}
\eeq

Using these identities and following the procedures 
outlined in \cite{MAP} it is possible after a suitable 
parameterization of the pole part in the $n$-particle 
formfactor to write down polynomial recursion equations 
for the $n$-particle deformed formfactors. For the 
Sinh-Gordon case first solutions have been found in 
\cite{MAX1}. Obviously, the limit $\beta \to 2 \pi$ 
reproduces the standard results for all objects 
introduced in this section 
\cite{MAP,OOTA}.

\bigskip
\bigskip

\bigskip
\bigskip
\bigskip

%%%%%%%%%%%%%%%%%%%%%%%%%%%%%%%%%%%%%%%%%%%%%%%%%%%%%%%%%%%%%%%%%%%%%

{\appendix{{\bf{Appendix: On the multiple sine-function}}}}

\bigskip

The multi-periodic sine-function was defined by (\ref{sinus-def}). 
Using the integral representation (\ref{gamma-def}) for $\Gamma_r$, 
we can write down an integral representation for $S_r$. 

\beq
\log S_r(x | {\underline{\omega}} ) = { {1} \over {2^{r+1} \pi i}} 
\int_{C_H} { {{\rm{d}} t}\over{t}} 
 {    { \log (-t) } \over {\prod_{i=1}^r 
      \sinh \left( { {\omega_i t}\over{2}} \right ) }} 
\left( (-1)^r {\rm{e}}^{xt - |{\underline{\omega}}| t/2} - 
            {\rm{e}}^{- xt + |{\underline{\omega}}| t/2} \right).
\label{sinus-int}
\eeq

It is of course possible using standard methods to rewrite 
this complex line integral as an integral over the positive 
real line. 

The function $S_r$ can be expanded in various ways. Usually, 
in the literature an infinite product expansion involving 
$\Gamma$-functions can be seen as minimal form factors. 
However, many of them can be easily shown not to be convergent. 

The correct expansion of $S_2$ in terms of standard $\Gamma$ 
functions is

\beq
\beqcol
S_2(x | \omega_1, \omega_2 ) & = &
{\rm{e}}^{  (\omega_1+ \omega_2 -2 x) ( \gamma+ \log ( 2 \pi ) 
+ 2 \log (\omega_1 / \omega_2 ) )/ ( 2 \omega_1) } 
{ {  \Gamma ( ( \omega_1+ \omega_2 -x)/\omega_1) } \over 
  {  \Gamma ( x/\omega_1) } }  \\
 & & \prod_{n=1}^{\infty}\left(   
  { {  \Gamma ( ( \omega_1+ \omega_2 -x + n \omega_2)/\omega_1) } \over 
  {  \Gamma ( (x + n \omega_2 )  /\omega_1) }}
 {\rm{e}}^{ - (\omega_1+\omega_2 -2 x)/(2 n \omega_1) } 
  ( n \omega_1 /\omega_2 )^{ - (\omega_1+\omega_2 -2 x)/\omega_1}.
\right) 
\eeqcol
\label{sin-prod-1}
\eeq

It may not be apparent, but this expression is symmetric in 
the two periods of $S_2$.

Another product expansion for $S_2$ was given in \cite{SHINTANI}.     

\beq
S_2(x|\omega_1,\omega_2)= \sqrt{i}\; {\rm{e}}^{ { { \pi i}\over{12} } 
( {{\omega_2} \over{\omega_1}} + {{\omega_1}\over {\omega_2 }} ) }
  {   { \prod_{n=0}^{\infty} 
      ( 1-q^n {\rm{e}}^{ { {2\pi i x}\over{\omega_1}}} )} \over
      {\prod_{n=1}^{\infty} 
   (1-q^{\prime \; n}  {\rm{e}}^{ { {2\pi i x}\over{\omega_2}}})}} 
{\rm{e}}^{ { { \pi i }\over { 2 \omega_1 \omega_2}} 
   ( x^2 - (\omega_1+ \omega_2) x)},
\label{sin-prod-2} 
\eeq

where $q=\exp (2 \pi i \omega_2/ \omega_1)$ and $q^{\prime} = 
\exp ( - 2 \pi i \omega_1 / \omega_2 )$.

\bigskip
\bigskip

{\bf{Acknowledgement}}

\medskip

This work was supported by EPSRC (grant GR/L26216).

\end{document}